\documentclass[twocolumn,superscriptaddress,preprintnumbers,showpacs,tightenlines,notitlepage]{revtex4}
\usepackage[latin1]{inputenc}
\usepackage{graphicx}
\usepackage{amssymb}
\usepackage{color}
\usepackage{float}
\usepackage{amsmath}
\usepackage{amsfonts}
\usepackage{dcolumn}
\usepackage{hyperref}
\usepackage{amsthm}
\usepackage{enumerate}
\usepackage{latexsym}
\usepackage{epsfig}
\usepackage{graphicx,psfrag,subfigure}
\usepackage{rotating}
\usepackage{tabularx}
\usepackage{epstopdf}
 \newcommand{\be}{\begin{equation}} \newcommand{\ee}{\end{equation}}
\newcommand{\bea}{\begin{eqnarray}} \newcommand{\eea}{\end{eqnarray}}
\newcommand{\bse}{\begin{subequations}} \newcommand{\ese}{\end{subequations}}

\newtheorem*{ques*}{Question}
\newtheorem{prop}{Proposition}
\newtheorem{thm}{Theorem}
\begin{document}

\title{Is cosmic censorship restored in higher dimensions?} 
\author{M. D. Mkenyeleye} 
\email{mkenyeleye@yahoo.co.uk\\Permanent Address:School of Mathematical Sciences, University of Dodoma, Tanzania.}
\affiliation{Astrophysics and Cosmology Research Unit, School of Mathematics, Statistics and Computer Science, University of KwaZulu-Natal, Private Bag X54001, Durban 4000, South Africa.}
\author{Rituparno Goswami}
\email{Goswami@ukzn.ac.za}
\affiliation{Astrophysics and Cosmology Research Unit, School of Mathematics, Statistics and Computer Science, University of KwaZulu-Natal, Private Bag X54001, Durban 4000, South Africa.}
\author{Sunil D. Maharaj}
\email{Maharaj@ukzn.ac.za}
\affiliation{Astrophysics and Cosmology Research Unit, School of Mathematics, Statistics and Computer Science, University of KwaZulu-Natal, Private Bag X54001, Durban 4000, South Africa.}
\begin{abstract}
In this paper we extend the analysis of gravitational collapse of spherically symmetric generalised Vaidya spacetimes to higher dimensions, in the context of the Cosmic Censorship Conjecture. We present the sufficient conditions on the generalised Vaidya mass function, that will generate a locally naked singular end state. Our analysis here generalises all the earlier works on collapsing higher dimensional generalised Vaidya spacetimes. With specific examples, we show the existence of classes of mass functions that lead to a naked singularity in four dimensions, which gets covered on transition to higher dimensions. Hence for these classes of mass function Cosmic Censorship gets restored in higher dimensions and the transition to higher dimensions restricts the set of initial data that results in a naked singularity.
\end{abstract}
\maketitle

\section{Introduction\label{Intro}}

The singularity theorems predict the occurrence of spacetime singularities for a wide class of theories of gravity under very generic conditions, namely the attractive nature of gravity, existence of closed trapped surfaces and no violations of causality in the spacetime \cite{HE}.  However these theorems do not say anything about the causal nature of these singularities, that is, if it is possible for future directed null geodesics from the close vicinity of these singular points, to escape to infinity. To avoid such scenarios where a  {\em naked singularity} exists that can causally influence the future infinities, the Cosmic Censorship Conjecture (CCC) was proposed by Penrose \cite{CCC}. This states that spacetime singularities produced by  the gravitational collapse of physically realistic matter fields are always covered by trapped surfaces. Hence the final state of continual gravitational collapse always leads to a black hole, where the singularity is shielded from any external observer. 

Though the general proof of this conjecture still remains elusive, there are a number of important counter-examples that show otherwise. Investigations of spherically symmetric dynamical collapse models in General Relativity for large classes of matter fields, in four dimensional spacetimes, indicate that there exist sets of initial data of non-zero measure, at the epoch of the commencement of the collapse, that lead to the formation of a locally naked singularity. In these cases the trapped surfaces do not form early enough to shield the singularity (or the spacetime fireball) from external observers.  It is also shown in these studies that families of future outgoing non-spacelike geodesics emerge from such a naked singularity, providing a non-zero measure set of trajectories escaping away \cite{Joshibook2,Lemos,Stefan}. Though these counter examples are mainly presented in case of spherical symmetry (with a few exceptions of non-spherical models), they suffice to be relevant because if the censorship is one of key aspect of gravitation theory, it should not depend on symmetries of spacetime. 

\subsection { The question}

To avoid the unpleasantries of nudity, the obvious question that arises (influenced by higher dimensional and emergent theories of gravity - e.g string theory or braneworld models),  is as follows:\\
\begin{ques*}
Does the transition to higher dimensional spacetimes (with compact or non-compact extra dimensions) restricts the above mentioned set of initial data that leads to a naked singularity?
\end{ques*}
In other words, how does the number of spacetime dimensions dictate the dynamics of trapped regions in the spacetime? This question is important as most of the proofs of the key theorems of black hole dynamics and thermodynamics demand the spacetimes to be future asymptotically simple, which is not possible if the censorship is violated \cite{HE}. If the locally naked singularities in 4-dimensional spacetime are naturally absent in higher dimensions, then that will be an argument in favour of higher dimensional (or emergent theories) of gravity, as in those cases the important results of black hole dynamics and thermodynamics would be more relevant. 

\subsection{ Earlier works}

To answer the above question, at least partially, one of the authors of the present paper showed the following important result \cite{RG1,RG2}: The naked singularities occurring in dust collapse from smooth initial data (which include those discovered by Eardley and Smarr \cite{Eardley}, Christodoulou \cite{Chris}, and Newman \cite{Newman}) are eliminated when we make transition to higher dimensional spacetimes. The cosmic censorship is then restored for dust collapse which will always produce a black hole as the collapse end state for dimensions $D\ge6$, under conditions such as the smoothness of initial data from which the collapse develops, which follows from physical grounds.

The physical reason behind the above result is that higher dimensional spacetimes favour trapped surface formation and the formation of horizons advance in time. Hence for dimensions greater than five, the vicinity of the singularity always gets trapped even before the singularity is formed, and hence the singularity is causally cut-off from any external observer.

Several other works on higher dimensional radiation collapse and perfect fluid collapse has been done \cite{ghosh1, ghosh2, ghosh3, ghosh4, Ghosh_2001, Kishor_2002, Beesham}, where the matter field is taken to be of a specific form (for example: perfect fluids with linear equation of state, pure radiation, charged radiation etc.). All of these studies give an indication that higher dimensions do favour trapping and hence the epoch of trapped surface formation advances as we go to higher dimensions.

\subsection {The present paper}

The main criticism of the dustlike models or pure perfect fluid models is that they are far too idealised. For any realistic massive astrophysical body, which is undergoing gravitational collapse, the pressure and the radiative processes must play an important role together. One of the known spacetimes that can closely mimic such a collapse scenario is the generalised Vaidya spacetime, where the matter field is a specific combination of {\em Type I} matter (whose energy momentum tensor has one timelike and three spacelike eigenvectors) that moves along timelike trajectories and {\em Type II} matter (whose energy momentum tensor has double null eigenvectors) that moves along null trajectories. Thus, a collapsing generalised Vaidya spacetime depicts the collapse of usual perfect fluid combined with radiation. Therefore the collapse scenario here is much closer  to what is expected for the collapse of a realistic astrophysical star. In our earlier paper \cite{Mkenyeleye_2014}, we investigated the gravitational collapse of generalised Vaidya spacetimes in four dimensions and developed  a general mathematical framework to study the conditions on the mass function such that future directed non-spacelike geodesics can terminate at the singularity in the past. In this paper:
\begin{enumerate}
\item We extend the earlier results to any arbitrary $N$-dimensional spacetimes. Though the general mathematical framework remains similar, the conditions on the mass function and it's derivatives for the collapse leading to a locally naked singularity, change as we make a transition to higher dimensional spacetimes.
\item Using explicit examples we show that there exist classes of mass functions, that lead the collapsing star to a naked singularity in four dimensions, will necessarily end in a black hole end state in dimensions greater than four. The reason for this remains the same as in dust models: formation of trapped surfaces is favoured in higher dimensions, and hence the vicinity of the central singularity gets trapped even before the singularity is formed. This gives a definite indication that the dynamics of trapped regions do depend on the spacetime dimensions for a large class of matter fields and the occurrence of trapped surfaces advance in time in higher dimensions.
\end{enumerate}
Unless otherwise specified, we use natural units ($c=8\pi G=1$) throughout this paper, Latin indices run from 0 to $N-1$. 
The symbol $\nabla$ represents the usual covariant derivative and $\partial$ corresponds to partial differentiation. 
We use the $(-,+,+,+,+,\cdots)$ signature and the Ricci tensor is obtained by contracting the {\em first} and the {\em third} indices of the Riemann tensor 
\begin{equation}
R^{a}{}_{bcd}=\Gamma^a{}_{bd,c}-\Gamma^a{}_{bc,d}+ \Gamma^e{}_{bd}\Gamma^a{}_{ce}-\Gamma^e{}_{bc}\Gamma^a{}_{de}\;,
\end{equation}
The Hilbert--Einstein action in the presence of matter is given by
\begin{equation}
{\cal S}=\frac12\int d^4x \sqrt{-g}\left[R-2\Lambda-2{\cal L}_m \right]\;,
\end{equation}
variation of which gives the Einstein field equations as
\be
G_{ab}+\Lambda g_{ab}=T_{ab}\;.
\ee

\section{Higher dimensional Generalised Vaidya spacetime \label{sec:two}}
The spherically symmetric line element for an $N$-dimensional generalised Vaidya spacetime is given as
\begin{equation}
ds^2 = -\left(1-\frac{2m(v,r)}{r^{(N - 3)}}\right)dv^2+2dvdr +r^2d\Omega_{(N-2)}^2\label{line-element},
\end{equation}
where
\begin{equation}
d\Omega_{(N-2)}^2 = \sum^{N-2}_{i=1}\left[\prod^{i-1}_{j=1}\sin^2(\theta^j)\right](d\theta^i)^2,
\end{equation}
is the metric on the $(N - 2)$ sphere in polar coordinates with $\theta^i$ being spherical coordinates. $m(v,r)$ is the generalized mass function related to the gravitational energy within a given radius $r$ \cite{Lake}, which can be carefully defined so that the energy conditions are satisfied. The coordinate $v$ represents the Eddington advanced time where $r$ is decreasing towards the future along a ray $v=Const.$ (ingoing). When $N = 4$, the line element reduces to the generalized Vaidya solution \cite{Wang} in $4$-dimensions.

Defining the following quantities
\begin{equation*}
\dot{m}(v,r)\equiv \frac{\partial m(v,r)}{\partial v},  \: \quad m'(v,r)\equiv  \frac{\partial m(v,r)}{\partial r},\,
\end{equation*}
we can write the non-vanishing components of the Ricci tensor as
\begin{subequations}\label{Ricci}
\begin{eqnarray}
%\nonumber to remove numbering (before each equation)
R^v_v &=& R^r_r=\frac{m''(v,r)}{r^{(N-3)}} - \frac{(N-4)m'(v,r)}{r^{(N-2)}}\label{Ricci1},\\
R^{\theta^1}_{\theta^1} &=& R^{\theta^2}_{\theta^2} = \cdots = R^{\theta^{(N-2)}}_{\theta^{(N-2)}} = \frac{2m'(v,r)}{r^{(N-2)}}.\label{Ricci2}
\end{eqnarray}
\end{subequations}\label{Einstein}
The Ricci scalar is given by
\begin{equation}\label{scalarcurvature}
 R = \frac{2m''(v,r)}{r^{(N-3)}} + \frac{4m'(v,r)}{r^{(N-2)}},
\end{equation}
while the non-vanishing components of the Einstein tensor are given by
\begin{subequations}\label{Einsteins:Eq}
\begin{eqnarray}
G^v_v &=& G^r_r=-\frac{(N-2)m'(v,r)}{r^{(N-2)}}\label{Einstein1},\\
G^r_v &=& \frac{(N-2)\dot{m} (v,r)}{r^{(N-2)}}\label{Einstein2},\\
G^{\theta^1}_{\theta^1} &=& G^{\theta^2}_{\theta^2}= \cdots = G^{\theta^{(N-2)}}_{\theta^{(N-2)}} = -\frac{m''(v,r)}{r^{(N-3)}}\label{Einstein3}.
\end{eqnarray}
\end{subequations}
The Energy Momentum Tensor (EMT) can be written in the form \cite{Husian_1996}
\begin{equation}
 T_{\mu\nu}=T^{(n)}_{\mu\nu}+T^{(m)}_{\mu\nu}, \label{EMT}
 \end{equation}
 where
 \begin{subequations}
 \begin{eqnarray}\label{EMT2}
 T^{(n)}_{\mu\nu}&=&\mu l_{\mu}l_{\nu},\\
 T^{(m)}_{\mu\nu}&=&(\rho+\varrho)(l_{\mu}n_{\nu}+l_{\nu}n_{\mu} )+ \varrho g_{\mu\nu}.
 \end{eqnarray}
 \end{subequations}
In the above,
 \begin{eqnarray}\label{const.}
 \mu &=& \frac{(N-2)\dot{m}(v,r)}{r^{(N - 2)}}, \quad \rho=\frac{(N-2)m'(v,r)}{r^{(N - 2)}}, \\ \nonumber
\varrho &=& -\frac{m''(v,r)}{r^{(N-3)}},
 \end{eqnarray}
 with $l_{\mu}$ and $n_{\mu}$ being two null vectors,
 \begin{eqnarray}
 l_{\mu} =\delta ^0_{\mu}, \quad  n_{\mu}=\frac{1}{2}\left [1-\frac{2m(v,r)}{r^{(N-3)}}\right]\delta ^0_{\mu}-\delta^1_{\mu},
 \end{eqnarray}
 where $l_{\mu}l^{\mu} =n_{\mu}n^{\mu}=0$ and $l_{\mu}n^{\mu}=-1$.

 Eq. \eqref{EMT} is taken as a generalized Energy-Momentum Tensor for the generalized Vaidya spacetime, with the component $T^{(n)}_{\mu\nu}$ being considered as the matter field that moves along the null hypersurfaces $v = \rm{constant}$, while
 $T^{(m)}_{\mu\nu}$ describes the matter moving along timelike trajectories.
 If the EMT of Eq. \eqref{EMT} is projected to the orthonormal basis, defined by the vectors,
 \begin{multline}
  E^{\mu}_{(0)} = \frac{l_{\mu} + n_{\mu}}{\sqrt{2}}, \quad E^{\mu}_{(1)} = \frac{l_{\mu} - n_{\mu}}{\sqrt{2}}, \quad E^{\mu}_{(2)} = \frac{1}{r}\delta^{\mu}_2,\\
 E^{\mu}_{(N)} = \frac{1}{r\sin\theta^1\sin\theta^2\sin\theta^3\cdots\sin\theta^{(N-2)}}\delta^{\mu}_N,
 \end{multline}
it can be found \cite{Wang}, that the symmetric EMT can be given as the $N\times N$ matrix,
\begin{equation}\label{EMTMatrix}
\left[T_{(\mu)(\nu)}\right] = \left[
                                 \begin{array}{ccccc}
                                   \frac{\mu}{2}+ \rho & \frac{\mu}{2} & 0 & \cdots & 0\\
                                   \frac{\mu}{2} & \frac{\mu}{2} - \rho & 0 & 0 & 0\\
                                   0 & 0 & \varrho  & 0 & 0\\
                                   \vdots & \cdots & 0 & \varrho  &\vdots\\
                                   0 & 0 & 0 & \cdots& \varrho \\
                                 \end{array}
                               \right].
 \end{equation}
 For this fluid the energy conditions are given as \cite{HE},  
 \begin{enumerate}
   \item \emph{The Weak and Strong energy conditions}:
   \begin{equation}\label{strongweakenergy}
    \mu\geq 0, \quad \rho\geq 0, \quad \varrho \geq 0, \quad (\mu\neq 0).
   \end{equation}
   \item \emph{The Dominant energy condition}:
   \begin{equation}\label{dominantenergy}
     \mu\geq 0, \quad \rho \geq \varrho \geq 0, \quad (\mu\neq 0).
   \end{equation}
 \end{enumerate}
 These energy conditions can be satisfied by suitable choices of the mass function $m(v,r)$.   

 \section{Higher dimensional collapse model \label{sec:3}}
In this section, we examine the gravitational collapse of a collapsing matter field in the generalized Vaidya spacetime when a spherically symmetric configuration of Type I and Type II matter collapses at the centre of symmetry in an otherwise empty universe which is asymptotically flat far away \cite{Joshi_1993}.

 If $K^{\mu}$ is the tangent to non-spacelike geodesics with $K^{\mu} = \frac{dx^{\mu}}{dk}$, where $k$ is the affine parameter, then $K^{\mu}_{;\nu}K^{\nu}=0$ and
 \begin{equation}
 g_{\mu\nu}K^{\mu}K^{\nu} =\beta , \label{tangents}
 \end{equation}
 where $\beta $ is a constant that characterizes different classes of geodesics with $\beta =0$ for null geodesic vectors, $\beta  < 0$ for timelike geodesics and $\beta  > 0$ for spacelike geodesics \cite{Joshi_1993}. Here we consider the case of null geodesics, that is,  $\beta =0$.

 We calculate the equations $dK^v/dk$  and $dK^r/dk$ using the Lagrange equations given by $L = \frac{1}{2}g_{\mu\nu}\frac{d{x}^{\mu}}{dk}\frac{d{x}^{\nu}}{dk}$ and Euler-Lagrange equations 
 \be
 \frac{\partial L}{\partial x^a}- \frac{d}{dk}\left(\frac{\partial L}{\partial{x^a_{,k}} }\right)= 0,
 \ee
In the case of the higher dimensional generalised Vaidya spacetime, these equations are given by
\begin{subequations}\label{tangents}
\begin{eqnarray}
\label{tangenteqn1}\frac{dK^v}{dk}+\left(\frac{(N-3)m(v,r)}{r^{(N-2)}} - \frac{m'(v,r)}{r^{(N-3)}}\right)\left(K^v\right)^2  &=& 0,\quad \quad \\
\frac{dK^r}{dk}+\frac{\dot{m}(v,r) }{r^{(N-3)}}\left(K^v\right)^2 &=& 0. \label{tangenteqn2}
\end{eqnarray}
\end{subequations}
All other components are considered to be $0$.
If we follow \cite{Dwivedi_1989} and write $K^v$ as
\begin{equation}\label{kvcomponent}
K^v=\frac{P(v,r)}{r},
\end{equation}
then using $K_{\mu}K^{\nu} = 0 $ we get
\begin{subequations}
\begin{eqnarray}
K^v &=& \frac{dv}{dk} = \frac{P(v,r)}{r}, \label{vcomponent} \\
K^r &=& \frac{dr}{dk} = \frac{P}{2r}\left(1 - \frac{2m(v,r)}{r^{(N-3)}}\right).\label{rcomponent}
\end{eqnarray}
\end{subequations}

\section{Conditions for Locally Naked Singularity\label{three}}
The nature (a locally naked singularity or a black hole) of the collapsing solutions can be characterized by the existence of radial null geodesics coming out of the singularity \cite{Joshi_1993, Ghosh_2001}.

The radial null geodesics of the line element \eqref{line-element} can be calculated using Eqs. \eqref{vcomponent} and \eqref{rcomponent}. These geodesics are given by the equation
\begin{equation}\label{nullgeodesics}
    \frac{dv}{dr} = \frac{2r^{(N-3)}}{r^{(N-3)} - 2m(v,r)}.
\end{equation}
This differential equation has a singularity at $r = 0, \quad v = 0$. Using the same techniques utilised in \cite{Tricomi_1961,Perko_1991,Mkenyeleye_2014}, Eq. \eqref{nullgeodesics} can be re-written near the singular point as
\begin{equation}\label{LimitvalueX0}
     \frac{dv}{dr} = \frac{2(N - 3)r^{(N-3)}}{(N - 3)r^{(N-3)} - 2m'_0r - 2\dot{m}_0v},
\end{equation}
where
\begin{subequations}
\begin{eqnarray}
m_0& =& \lim\limits_{v\to 0, r \to 0}m(v,r),\\ \dot{m}_0 &=& \lim\limits_{v\to 0, r \to 0}\frac{\partial}{\partial v}m(v,r), \\m'_0 &= &\lim\limits_{v\to 0, r \to 0}\frac{\partial}{\partial r}m(v,r)\, .
\end{eqnarray}
\end{subequations}
\subsection{Existence of outgoing nonspacelike geodesics}
We can clearly  see that Eq. \eqref{LimitvalueX0} has a singularity at $v = 0, \quad r = 0$. 
The classification of the tangents of both radial and non-radial outgoing non-spacelike geodesics terminating at the singularity in the past can be given by the limiting values at $v = 0, r = 0$. The conditions for the existence for such geodesics have been described in detail in \cite{Mkenyeleye_2014} using the concept of contraction mappings.
The existence of these radial null geodesics also characterizes the nature (a naked singularity or a black hole) of the collapsing solutions. If we let $X$ to be the limiting value at $r = 0, v = 0$, we can determine the nature of this limiting value on a singular geodesic as
 \begin{equation}\label{limitingvalue}
    X_0 = \lim_{\substack{v\rightarrow 0, r\rightarrow 0}}X = \lim_{\substack{v\rightarrow 0, r\rightarrow 0}}\frac{v}{r}.
 \end{equation}
Using a suitably chosen mass function, Eq. \eqref{LimitvalueX0} and l'Hopital's rule, we can explicitly find the expression for the tangent values $X_0$ which governs the behaviour of the null geodesics near the singular point. Thus, the nature of the singularity can then be determined by studying the solution of this algebraic equation. This expression can be calculated as
\begin{eqnarray}\label{limitingvalueX0}
 X_0 & = &\lim_{\substack{v\rightarrow 0, r\rightarrow 0}}\frac{dv}{dr} \nonumber \\
 &=& \lim_{\substack{v\rightarrow 0, r\rightarrow 0}}\frac{2(N - 3)r^{(N-4)}}{(N - 3)r^{(N-4)} - 2m'_0 - 2\dot{m}_0 X_0}.
 \label{X_0 value}
\end{eqnarray}
\subsection{Apparent Horizon}
The existence of the apparent horizon, which is the boundary of the trapped surface region in the spacetime also determines the nature of the singularity. If at least one value of the limiting positive values $X_0$ is less than the slope of the apparent horizon at the central singularity, then the central singularity is locally naked with the outgoing radial null geodesics escaping from the past to the future.
 
For the generalized higher dimensional Vaidya spacetime, the apparent horizon is defined by
\begin{equation}\label{Eq:Apparenthorizon}
2m(v,r) = r^{(N-3)}.
\end{equation}
The slope of the apparent horizon can be calculated as follows:
\begin{subequations}
\begin{equation}
2\frac{dm(v,r)}{dr} = (N-3)r^{(N-4)},
\end{equation}
\begin{equation}
2\left(\frac{\partial m}{\partial v}\right)\left(\frac{dv}{dr}\right)_{AH} + 2\frac{\partial m}{\partial r} = (N-3)r^{(N-4)}.
\end{equation}
\end{subequations}
Thus, the slope of the apparent horizon at the central singularity is given by
\begin{equation}
X_{AH} = \left(\frac{dv}{dr}\right)_{AH} = \lim\limits_{v\to 0, r \to 0}\frac{(N-3)r^{(N-4)}-2m'_0}{2\dot{m}_0},
\label{XAH}
\end{equation}

\begin{table*}
\caption{ Algebraic equations for $X_0$ and $X_{AH}$ for different values of $n$ and $N$}
\label{Table: X0expressions}
\begin{ruledtabular}
\begin{tabular}{lll}
    $n$ and $N$ & Expression for $X_{0}$ & Expression for $X_{AH}$\\\hline
 $n = 1$, $N=4$ & $\lambda_1 X_0^2 - X_0 +2 = 0$, & $X_{AH} = \frac{1}{\lambda_1}$ \\ \hline
 $n =2, N=4$ & $\lambda_2X_0^3 -\lambda_1 X_0^2 + X_0 - 2 =0$ & $\lambda_2X_{AH}^2 - \lambda_1 X_{AH} + 1 = 0$\\ \hline
 $n=2$, $N=5$ & $2\lambda_2X_0^5 - 3\lambda_1 X_0^4 + 2X_0 - 4 = 0$ &$2\lambda_2X_{AH}^4 - 3\lambda_1X_{AH}^3 + 2 =0 $\\ 
 \end{tabular}
 \end{ruledtabular}
\end{table*}

\subsection{Sufficient conditions}

We can now write the sufficient conditions for the existence of a locally naked central singularity for a collapsing generalised Vaidya spacetime in arbitrary dimensions $N$, which we state in the following proposition:
\begin{prop}
Consider a collapsing N-dimensional generalised Vaidya spacetime from a regular epoch, with a mass function $m(v,r)$, that obeys all physically reasonable energy conditions and is differentiable in the entire spacetime. If the following conditions are satisfied :
\begin{enumerate}
\item The limits of the partial derivatives of the mass function $m(v,r)$ exist at the central singularity,
\item There exist one or more positive real roots $X_0$ of the equation (\ref{X_0 value}),
\item At least one of the positive real roots of $X_0$ is less than the smallest root of equation (\ref{XAH}),
\end{enumerate}
then the central singularity is locally naked with outgoing $C^1$ radial null geodesics escaping to the future.
\end{prop}

We emphasise here, that all the previous works of higher dimensional generalised Vaidya collapse \cite{Ghosh_2001, Kishor_2002, Beesham} are special cases of the general analysis presented above. In the next section, we give a specific example to transparently demonstrate the effect of transition to higher dimensions on the nature of the central singularity.

\section{A general Laurent expandable mass function\label{general}}
\begin{table*}
\caption{ Values of $X_0$ and ${\rm Min}[X_{AH}]$ for different dimensions for $\lambda_1 = 5.0$, $\lambda_2= 0.01$, $\lambda_3 = 2.3$, $\lambda_4 = 0.05$.}
\label{Table:2}
\begin{ruledtabular}
\begin{tabular}{lll}
 $N$ &  $X_{0}$ &  ${\rm Min}[X_{AH}]$\\\hline
 $4$ &  $0.204$     &  $1.472$ \\ \hline
 $5$ &  $1.806$     &  $0.526$\\ \hline
 $6$ &  $1.902$     &  $0.672$\\ \hline
 $7$ &  $1.948$     &  $0.751$\\ 
 \end{tabular}
 \end{ruledtabular}
 \end{table*}

We consider here a Laurent expandable mass function of the generalized Vaidya spacetime in higher dimensions in the general form as 
\begin{equation}\label{eq:general_mass}
2m(v,r) = \lambda_1 m_1(v) - \lambda_2\frac{m_2(v)}{r^{(N-3)}} -\lambda_3\frac{m_3(v)}{r^{(N-2)}} + \cdots,
\end{equation}
where
\begin{equation*}
m_n(v) = v^{(2N+n-8)}, \: n = 1,2, \cdots \: \: \:  \text{ and $\lambda_n$'s are constants}. 
\end{equation*}
Using Eq. \eqref{limitingvalueX0} and \eqref{XAH}, we get the expression of the tangent to null geodesics $X_0$ and tangent to the apparent horizon $X_{AH}$ in higher dimensions as
\begin{widetext}
\begin{equation}\label{eqn:limitgeneral}
X_0 = \frac{2(N-3)}{(N-3)-(2N-7)\lambda_1 X_0^{(2N-7)}+ (N-3)\left(\lambda_2X_0^{2N-6}+\lambda_3 X_0^{(2N-5)}+\lambda_4 X_0^{(2N-4)}+\cdots\right)}.
\end{equation}
\begin{equation}\label{Eq:apparentgeneral}
X_{AH} = \frac{(N-3) - (N-3)\lambda_2 X_{AH}^{(2N-6)}-(N-2)\lambda_3X_{AH}^{(2N-5)}- (N-1)\lambda_4X_{AH}^{(2N-4)}- \cdots}{\lambda_1(2N-7)X_{AH}^{(2N-8)}-(2N-6)\lambda_2X_{AH}^{(2N-7)}-(2N-5)\lambda_3X_{AH}^{(2N-6)} - \cdots},
\end{equation}
\end{widetext}
These expressions can be written in the general form as 
\begin{equation}\label{eq:general_x0}
\sum\limits^{\infty}_{n=1}\left(f_{n}(N,\lambda_i)X_{0}^{(2N+n-7)}\right) + (N-3)X_{0} - 2(N-3) = 0, 
\end{equation}
and
\begin{equation}
\sum\limits^{\infty}_{n=1}g_n(N,\lambda_i)X_{AH}^{(2N+n-8)} - (N-3) = 0.
\end{equation}
where $f_{n}(N,\lambda_i)$ and $g_n(N,\lambda_i)$ are some functions of $N$ and the $\lambda_i$'s.

These expressions can explicitly be solved for $X_0$ and $X_{AH}$ using some specific values of $n$, $N$ and $\lambda_i$'s (see Table \ref{Table: X0expressions}) and then one can make conclusions about the nature of the singularity by using the following conditions:
\begin{enumerate}[(i)]
\item If there is no positive real solution for $X_0$, then there are no outgoing null geodesics from the singularity and the singularity is causally cut off from the external observer.
\item If there is no real solution for $X_{AH}$, then there are no trapped surfaces and the singularity is globally naked, provided there is at least one positive real root of $X_0$.
\item If there are one or multiple real solutions for $X_{AH}$ with the smallest solution less than $X_0$, then it can be concluded that the collapse results in a black hole end state.
\item If the smallest solution ${\rm{Min}}[X_{AH}] $ is greater than any one of the positive solutions of $X_0$, then there will be future directed null geodesics from the singularity and hence the singularity is locally naked.
\end{enumerate}
We can easily see from Table \ref{Table: X0expressions}, that the general expression obtained here contains the expressions for $X_0$ and $X_{AH}$ corresponding to Vaidya collapse in $4$-D ($n =1, N=4$) \cite{Dwivedi_1989,Joshi_1993}, charged Vaidya-de Sitter in $4$-D ($n=2, N=4$) \cite{Beesham} and charged Vaidya in $5$-D ($n=2, N=5$) \cite{Kishor_2002}.

 \begin{table*}
\caption{Range for $X_0$ and ${\rm Min}[X_{AH}]$ for different dimensions: $\{4.8<\lambda_1< 5.25, 0.009<\lambda_2 < 0.012, 2.25 <\lambda_3<2.38, \lambda_4 = 0.05\}$}
\label{Table:3}
\begin{ruledtabular}
\begin{tabular}{lll}
 $N$ & Range for $X_{0}$ & Range for ${\rm Min}[X_{AH}]$\\\hline
 $4$ &  $0.194 <X_{0}< 0.213$&  $1.377 < {\rm Min}[X_{AH}]<1.483$ \\ \hline
 $5$ &  $1.789<X_{0}<1.818$ &  $0.517 < {\rm Min}[X_{AH}]< 0.534$\\ \hline
 $6$ &  $1.884 <X_{0}<1.915$ &  $0.665 < {\rm Min}[X_{AH}]<0.678$\\ \hline
 $7$ &  $1.930 <X_{0}<1.962$ &  $0.745 < {\rm Min}[X_{AH}]< 0.756$\\ 
 \end{tabular}
 \end{ruledtabular}
 \end{table*}
\begin{table*}
\caption{Range for $X_0$ and ${\rm Min}[X_{AH}]$ for different dimensions: $\{\lambda_1 =5.2, 0.009<\lambda_2 < 0.012,\lambda_3=2.3,0\leq\lambda_4 <0.4\}$}
\label{Table:4}
\begin{ruledtabular}
\begin{tabular}{lll}
 $N$ & Range for $X_{0}$ & Range for ${\rm Min}[X_{AH}]$\\\hline
 $4$ &  $1.3934 <X_{0}<1.3941$& $1.5010 < {\rm Min}[X_{AH}]<1.5017$ \\ \hline
 $5$ &  $1.8406<X_{0}<1.8412$ &  $0.5184 < {\rm Min}[X_{AH}]< 0.5185$\\ \hline
 $6$ &  $1.9387 <X_{0}<1.9393$ &  $0.6658 < {\rm Min}[X_{AH}]<0.6659$\\ \hline
 $7$ &  $1.9865 <X_{0}<1.9872$ &  $0.7431 < {\rm Min}[X_{AH}]< 0.7432$\\ 
 \end{tabular}
 \end{ruledtabular}
 \end{table*}

\subsection{Example: Class of naked singularity in 4D being eliminated in higher dimensions}

In this section we will consider a specific example, that can be easily generalised to an open set, to show explicitly how a naked singularity in four dimensions gets covered in 
higher dimensions. Let us consider a scenario where $n=4$.
In this case the expression for $X_0$ and $X_{AH}$ become
 \begin{multline}\label{Eq: X_0n4}
(2N-7)\lambda_1X_0^{2N-6} -(N-3)\lambda_2X_0^{2N-5}-(N-3)\lambda_3X_0^{2N-4}\\-(N-3)\lambda_4X_0^{2N-3} - (N-3)X_0+2(N-3) = 0,
\end{multline}
and 
\begin{multline}\label{Eq: X_AHn4}
(2N-7)\lambda_1X_{AH}^{2N-7} -(N-3)\lambda_2X_{AH}^{2N-6}-(N-3)\lambda_3X_{AH}^{2N-5}\\-(N-3)\lambda_4X_{AH}^{2N-4}-(N-3) = 0.
\end{multline}

We can solve these equations numerically to get the values of $X_0$ and $X_{AH}$ in different dimensions. For our calculations we took 
$\lambda_1 = 5.0$, $\lambda_2= 0.01$, $\lambda_3 = 2.3$, $\lambda_4 = 0.05$. From Table \ref{Table:2} we can easily see that in 4 dimensions, this class of mass function leads to a naked singularity, as the trapped surfaces do not form early enough to shield the singularity from outside observers. However when we make the transition to higher dimensions we see that the value of the tangent to the outgoing null geodesic from the central singularity is greater than the slope of the apparent horizon curve at the central singularity. In this case the outgoing null direction is within the trapped region and hence the singularity is causally cut off from the external observer. By the continuity of the mass function considered above, this can be easily converted to a open set in the mass function space, where this scenario continues to be true and we shall explicitly prove this in the following subsection.

\subsection{Proof of existence of open set of mass functions with the above properties}

Having found out a specific example of a mass function for which the naked singularities in 4D is eliminated when we go to higher dimension, we would now require to prove that such a mass function is generic in the sense that there exists a open set of such mass functions in the function space. Since this problem of deducing the nature of the central singularity is reduced to finding and comparing real roots of polynomials (\ref{Eq: X_0n4}) and (\ref{Eq: X_AHn4}), all we need to show here is the real roots of these polynomials are continuous functions of the coefficients. 

To do this, first of all we observe that the roots that are given in the Table \ref{Table:2} are all of multiplicity one. This can be easily seen by differentiating the LHS of (\ref{Eq: X_0n4}) and (\ref{Eq: X_AHn4}) and substituting the roots to find non-zero values. Now, for any complex polynomial $p(z)$ of degree $n\ge1$ with $m$ distinct roots $\{\alpha_1,\cdots, \alpha_m\}$, ($1\le m\le n$), let us define the quantity $R_0(p)$ as follows:
\begin{equation}
  R_0(p)=\begin{cases}
    \frac12, & \text{if $m=1$}.\\
    \frac12 {\rm min}|\alpha_i-\alpha_j|, i\le j\le m, & \text{if $m>1$}.
  \end{cases}
\end{equation}
We now state the well known result of complex analysis \cite{alen}:
\begin{thm}
Let $p(z)$ be a polynomial of degree $n\ge1$, with real coefficients $\{\mu_k\}$. Suppose $\alpha$ be a real root of $p(z)$ of multiplicity one. Then for any $\epsilon$ with $0\le\epsilon\le R_0(p)$, there exists a $\delta(\epsilon)>0$ such that any polynomial $q(z)$ with real coefficients ${\nu_k}$ and $|\mu_k-\nu_k |\le \delta$, has a real root $\beta$ with 
$|\alpha-\beta|\le \epsilon$.
\end{thm}
The above theorem shows that if a polynomial $p(z)$ with real coefficients has a real root
$\alpha$ of multiplicity one, then any polynomial $q(z)$ obtained by small (real) perturbations to
the coefficients of $p(z)$ will also have a real root in a neighbourhood of $\alpha$. That is, not only
the root depends continuously on coefficients, but also it remains real, under
sufficiently small perturbations of coefficients.

This results directly translates to our problem of open set of mass functions in the mass function space. Once we have a specific example as shown in Table \ref{Table:2}, any perturbations around that will have the same outcome as far as the nature of the singularities are concerned. Hence this class of mass functions is not fine tuned, but quite generic and the outcome is stable under perturbations. 

\subsection{Numerical verification}

We would now like to verify explicitly, with the aid of numerical calculations, the results in the previous subsection. We numerically solve equations (\ref{Eq: X_0n4}) and (\ref{Eq: X_AHn4}) to get the values of $X_0$ and $X_{AH}$ in different dimensions to show that there exists a set of parameter intervals for which the mass function leads to a naked singularity in 4 dimensions and a black hole in higher dimensions. For example, some of the intervals are $\{4.8<\lambda_1< 5.25, 0.009<\lambda_2 < 0.012, 2.25 <\lambda_3<2.38, \lambda_4 = 0.05\}$ with the range values shown in Table \ref{Table:3} and $\{\lambda_1 = 5.2, 0.009<\lambda_2 <0.012, \lambda_3 = 2.3, 0\leq\lambda_4<0.4\}$ as shown in Table \ref{Table:4}, we can easily see that in 4 dimensions, these classes of mass function leads to a naked singularity, as the trapped surfaces do not form early enough. However when we make the transition to higher dimensions, the final outcome is a black hole.

\subsection{The result}

As a result of our detailed analytical and numerical investigations of the previous subsections, we can state the following proposition:
\begin{prop}
There exist classes of mass function in generalised Vaidya spacetimes, that produce a locally naked central singularity in 4 dimensions, but these naked singularity gets eliminated in higher dimensions due to temporal advancement of trapped surface formation.
\end{prop}

 \section{Summing it  all up\label{conclusion}}
In this paper we extended our analysis of the gravitational collapse of generalised Vaidya spacetime in 4-dimensions \cite{ Mkenyeleye_2014}, to spacetimes of arbitrary dimensions, in the context of the Cosmic Censorship Conjecture. We found the sufficient conditions on the generalised Vaidya mass function, that generates a locally naked central singularity that can causally communicate with an external observer. We carefully investigated the effect of the number of dimensions on the dynamics of the trapped regions, by studying the slope of the apparent horizon curve at the central singularity. 

By considering specific examples, we showed that there exist classes of mass functions for which a naked singularity in 4-dimensions gets covered as we make the transition to higher dimensional spacetime. Interestingly, the reason for this is same as in the case of dust collapse. From our analysis here, we can easily see that for a wide class of matter fields, a transition to higher dimensions favours trapped surface formation and the epoch of trapping advances as we go to higher dimensions. This makes the vicinity of the central singularity trapped even before the singularity is formed, and hence it is necessarily covered. 

Therefore, we can safely conclude that for a large class of matter fields, which include both Type I and Type II matter, transition to higher dimensions does indeed restrict the set of physically realistic initial data, that leads to the formation of a locally naked singularity.

\section {Acknowledgements \label{acknowledge}}
We are indebted to the National Research Foundation and the University of KwaZulu-Natal for financial support.
SDM acknowledges that this work is based upon research supported by the South African Research Chair Initiative of the
Department of Science and Technology. MDM extends his appreciation to the University of Dodoma in Tanzania for a study leave. We also wish to thank the anonymous referee for his constructive comments on this paper.


\thebibliography{}
\bibitem{HE}
S. W. Hawking and G. F. R.
Ellis, \emph{\it The Large Scale Structure of Spacetime},
Cambridge University Press, Cambridge,1973.

\bibitem{CCC}
R. Penrose, {\it Gravitational Collapse: The Role of General Relativity}, Riv. Nuovo Cimento, Num. Sp. I, 1969.

\bibitem{Joshibook2} 
 P. S. Joshi,{ \it Gravitational Collapse and Spacetime Singularities}, Cambridge University press, Cambridge, 2007.
 \bibitem{Lemos} J. P. S. Lemos, Phys. Rev. Lett. 68, 1447 (1992).
 \bibitem{Stefan} R. Baier, H. Nishimura, S. A. Stricker, arXiv:1410.3495 [gr-qc].
 
 \bibitem{RG1} R. Goswami and P. S. Joshi, Phys. Rev. D {\bf 69} 104002 (2004).
 \bibitem{RG2} R. Goswami and P. S. Joshi, Phys.Rev. D{\bf 69} 044002 (2004).
 
\bibitem{Eardley}  D. M. Eardley,. et al. Phys.Rev. D {\bf19} 2239 (1979).
\bibitem{Chris}  D. Christodoulou, Commun.Math.Phys. {\bf 93}  171 (1984).
\bibitem{Newman}  R. P. A. C. Newman. Class. Quant. Gravit. {\bf 3} 527 (1986).
\bibitem{ghosh1} S. G. Ghosh, A. K. Dawood, Gen. Rel. Gravit. {\bf 40}, 9-21, (2008).
\bibitem{ghosh2} S. G. Ghosh, D. W. Deshkar, Astrophys. Space Sci. {\bf 310} 111-117, (2007).
\bibitem{ghosh3} N. Dadhich, S .G. Ghosh, D. W. Deshkar, Int. J. Mod. Phys. A{\bf 20},1495-1502 (2005).
\bibitem{ghosh4} S. G. Ghosh, R. V. Saraykar,  Phys. Rev. D{\bf 62},107502, (2000).
\bibitem{Ghosh_2001} S. G. Ghosh and N. Dadhich, Phys. Rev. D \textbf{64}, 047501 (2001).
\bibitem{Kishor_2002} K. D. Patil, Pramana J. Physics {bf 60}, 423 (2003).
\bibitem{Beesham} A. Beesham, S.G. Ghosh, Int. J. Mod. Phys. D \textbf{12}, 801 (2003).

\bibitem{Mkenyeleye_2014} M. D. Mkenyeleye, R. Goswami and S. D. Maharaj, Phys. Rev. D \textbf{90}, 064034 (2014).



\bibitem{Lake} K. Lake and T. Zannias, Phys. Rev. D \textbf{43}, 1798 (1990).
\bibitem{Wang} A. Wang and Y. Wu, Gen. Rel. Gravit. \textbf{31} 107 (1999).
\bibitem{Husian_1996} V. Husain, Phys. Rev. \textbf{D53}, R1759 (1996).
 
\bibitem{Joshi_1993} P.S. Joshi, \emph{Global Aspects in Gravitation and Cosmology}, Clarendon press, Oxford (1993).
\bibitem{Dwivedi_1989} H. Dwivedi and P. S. Joshi, Classical. Quantum Gravity. \textbf{6} 1599 (1989).
 
\bibitem{Tricomi_1961} F. G. Tricomi, {\it Differential Equations} (Blackie \& Son Ltd., London, 1961).
\bibitem{Perko_1991} L. Perko, {\it Differential Equations and Dynamical Systems} (Springer-Verlag, New York, 1991).
\bibitem{alen} See for example http://users.ices.utexas.edu/alen/articles/, and the references therein.
\end{document}